\documentclass[lettersize,journal]{IEEEtran}
\usepackage{amsmath,amsfonts}
\usepackage{algorithmic}
\usepackage{algorithm}
\usepackage{array}
\usepackage[caption=false,font=normalsize,labelfont=sf,textfont=sf]{subfig}
\usepackage{textcomp}
\usepackage{stfloats}
\usepackage{url}
\usepackage{verbatim}
\usepackage{graphicx}
\usepackage{cite}
\usepackage{cite}
\usepackage{amsmath,amssymb,amsfonts}
\usepackage{algorithmic}
\usepackage{booktabs}
\usepackage{xcolor}
\usepackage{multirow}
\usepackage{multicol}
\usepackage{url}
\usepackage{makecell}
\usepackage{xcolor}
\usepackage{rotating}
\usepackage{nicematrix}
\usepackage{colortbl}
\definecolor{thisworkblue}{RGB}{230,241,252}
\usepackage{hhline}

\newcolumntype{C}[1]{>{\centering\arraybackslash}p{#1}}
\newcolumntype{L}[1]{>{\raggedright\arraybackslash}p{#1}}

\usepackage{etoolbox}

\usepackage{eso-pic}
\usepackage[hidelinks]{hyperref}
\newcommand{\arxivpublicationnotice}{%
\AddToShipoutPictureFG*{%
\AtPageUpperLeft{\put(\LenToUnit{0.65in},\LenToUnit{-9pt}){\makebox(0,0)[lt]{%
\parbox[t]{7.2in}{\fontsize{7.5}{8.5}\selectfont\raggedright
Author's accepted manuscript. Published in \textit{IEEE Solid-State Circuits Letters}, Early Access, 2026.\\
X. Chen, Y. Xu, Y. Hou, B. Liu, A. G. Richardson, and X. Liu, ``A Mixed-Precision CNN-LSTM Accelerator Featuring Long-Kernel Memory Reuse and Dynamic Quantization for Sleep Staging.''\\
DOI \href{https://doi.org/10.1109/LSSC.2026.3746976}{10.1109/LSSC.2026.3746976}. IEEE Xplore \url{https://ieeexplore.ieee.org/document/11753942}.
}}}}%
\AtPageLowerLeft{\put(\LenToUnit{0.65in},\LenToUnit{9pt}){\makebox(0,0)[lb]{%
\parbox[b]{7.2in}{\fontsize{7}{8}\selectfont\raggedright
\copyright\ 2026 IEEE. Permission from IEEE must be obtained for all other uses, in any current or future media, including reprinting/republishing this material for advertising or promotional purposes, creating new collective works, for resale or redistribution to servers or lists, or reuse of any copyrighted component of this work in other works.
}}}}%
}}

\begin{document}

\title{A Mixed-Precision CNN-LSTM Accelerator Featuring Long-Kernel Memory Reuse and Dynamic Quantization for Sleep Staging}

\author{Xinyu~Chen*, Yaqian~Xu*, Yuhan~Hou, Bowen~Liu, Andrew~G.~Richardson, and~Xilin~Liu
\thanks{*These authors contributed equally to the work.}
\thanks{Xinyu Chen, Yaqian Xu, Yuhan Hou, Bowen Liu, and Xilin Liu are with the Edward S. Rogers Sr. Department of Electrical and Computer Engineering, University of Toronto, Toronto, ON M5S 2E4, Canada.}
\thanks{Andrew~G.~Richardson is with the Department of Neurosurgery, University of Pennsylvania, Philadelphia, PA 19104, USA.}
}

\markboth{IEEE Solid-State Circuits Letters}{Chen \MakeLowercase{\textit{et al.}}: CNN-LSTM Accelerator}

\maketitle
\thispagestyle{empty}
\arxivpublicationnotice

\begin{abstract}
This letter presents a mixed-precision CNN-LSTM accelerator for EEG-based sleep staging. To efficiently execute long temporal convolutions, a serpentine dataflow propagates activations between neighboring processing elements, reducing memory accesses by 6.2$\times$ and 2.25$\times$ compared with output- and weight-stationary dataflows, respectively. A hybrid quantization scheme applies runtime dynamic quantization to the LSTM and MLP layers and assigns mixed precision on a layer-by-layer basis, reducing model size by 4.16$\times$. Combined with a bit-flexible MAC architecture and channel-grouped memory scheduling, the accelerator supports INT2/4/8/16 inference. Fabricated in 65\,nm CMOS, the accelerator consumes 0.40\,mW and achieves a 258$\times$ speedup over an iso-frequency 32b \mbox{RISC-V} processor. Evaluated across all 62 subjects in the public MASS-SS3 dataset using 31-fold leave-two-subject-out cross-validation, the mixed-precision approach with hybrid quantization maintains 5-class accuracy within 0.15\% of the floating-point baseline when optimized for general sleep staging and N3 sensitivity within 0.51\% when optimized for deep-sleep detection. The developed techniques are broadly applicable to EEG-based DL workloads.
\end{abstract}


\section{Introduction}
\begin{figure}[!t]
    \centering
    \includegraphics[width=.9\linewidth]{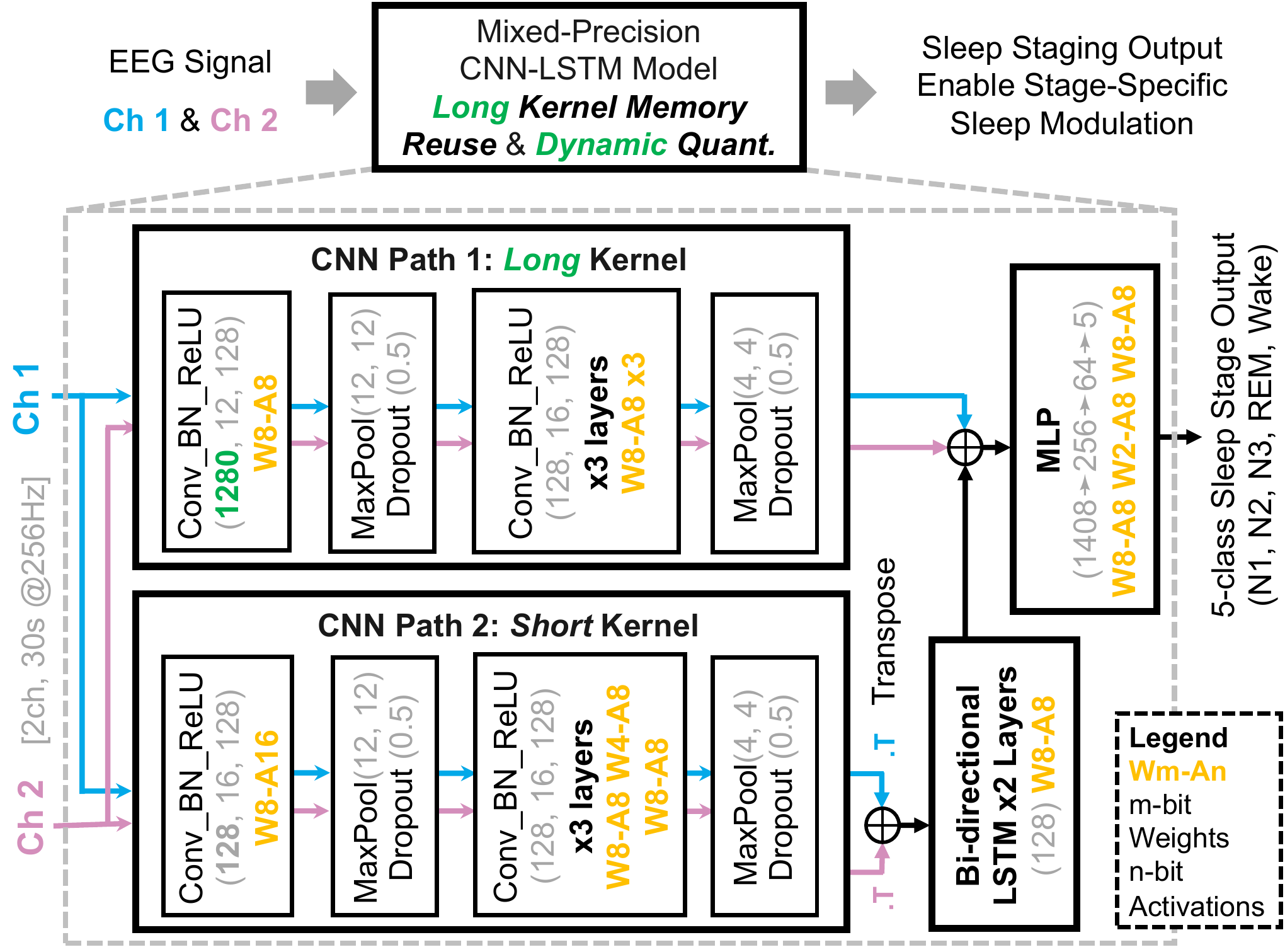} \vspace{-2mm}
    \caption{Overall operational principle and the developed dual-path hybrid CNN-LSTM model for sleep staging with key parameters highlighted.} \vspace{-4mm}
    \label{fig:intro_model}
\end{figure}

\IEEEPARstart{S}{leep} disorders affect nearly 40\% of adults, leading to cognitive decline and long-term health complications~\cite{wunderlin2020role}. Sleep-stage-specific auditory stimulation (Fig.~\ref{fig:intro_model}) can enhance sleep quality, but requires precise deep sleep (N3 stage) detection using advanced deep learning (DL) models~\cite{zhou2026fpga}.
Local DL inference on wearable EEG sensors reduces latency and wireless transmission energy, but the demanding DL workloads often exceed the available compute and power budgets, particularly when sub-milliwatt operation is needed to sustain overnight monitoring with small, lightweight batteries.
Furthermore, sleep-staging models require long convolutional kernels to capture slow EEG oscillations, causing excessive memory access and poor data reuse on conventional DL accelerators predominantly optimized for short kernels~\cite{chen2017eyeriss}. Moreover, error accumulation in recurrent networks causes conventional static quantization methods to fail, particularly in recurrent layers. Together, these bottlenecks highlight the critical need for an energy-efficient hardware accelerator tailored for sleep staging and many other EEG-based applications that share similar computational workloads.

This work addresses these challenges through the development of three design techniques. First, a serpentine dataflow is proposed to efficiently map long kernels to the processing element (PE) array through local memory reuse. Second, dynamic quantization with mixed precision is utilized to reduce model size while preserving inference accuracy. Third, a bit-flexible multiplier and memory scheduling scheme are developed to efficiently support mixed-precision computation. This design was initially introduced as part of a sleep modulation SoC in~\cite{xu2026inear}.
This letter focuses exclusively on the DL accelerator, providing detailed architectural analysis, an expanded evaluation across all subjects in the dataset, further optimization of the quantization strategy, and experimental characterization.

\section{DL Model and Accelerator Overview}

A hybrid CNN-LSTM model is developed and trained on the public MASS-SS3 dataset~\cite{o2014montreal} for 5-class sleep staging using two EEG channels near the forehead, directly relevant to forehead-based wearable sleep monitoring.
Each input comprises a 30\,s EEG epoch, with one sleep-stage prediction generated per epoch.
The model features two parallel, complementary coarse-fine feature-extraction CNN branches, followed by a bidirectional LSTM and a multilayer perceptron (MLP) layer (Fig.~\ref{fig:intro_model}). The coarse CNN branch employs long kernels of up to 1280 samples to extract low-frequency temporal features over a large receptive field, while the fine-grained branch uses shorter kernels to capture local waveform characteristics.

\begin{figure}[!t]
    \centering
    \includegraphics[width=1\linewidth]{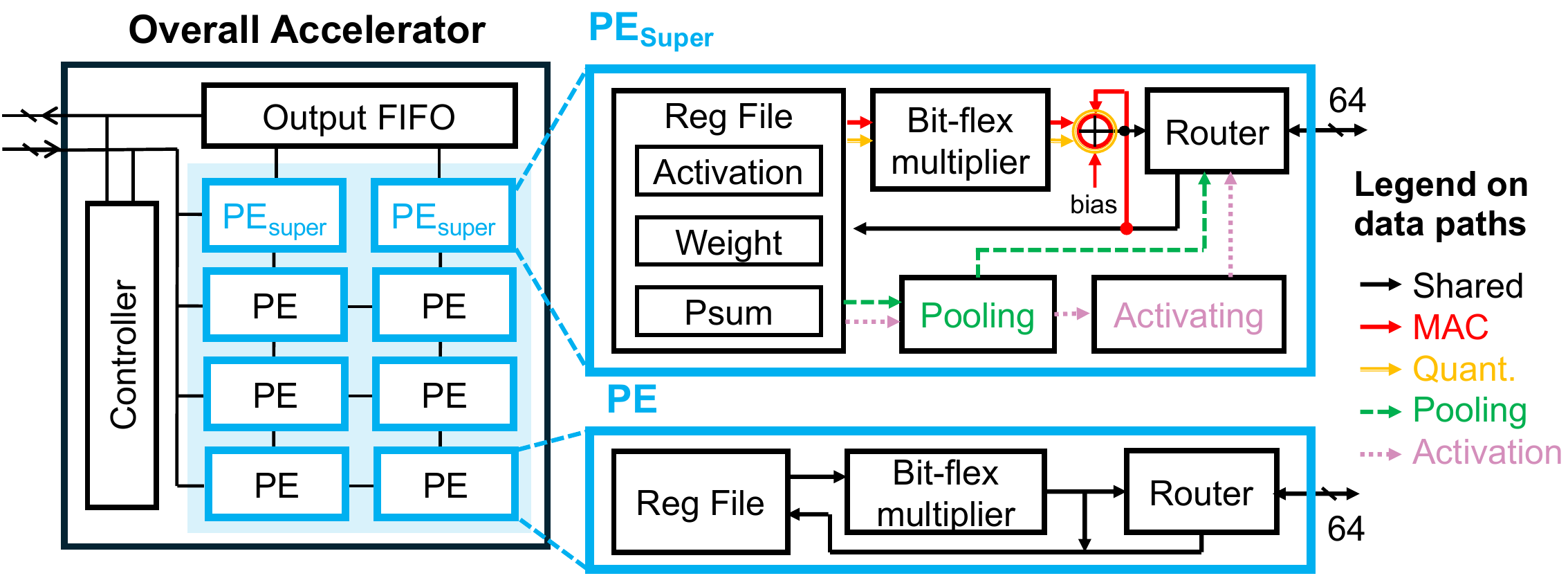} \vspace{-5pt} \vspace{-5mm}
    \caption{Overall architecture of the developed DL accelerator and data paths of the Super PE in different operation modes.} \vspace{-4mm}
    \label{fig:architecture}
\end{figure}

The accelerator has a 1\,mm$^2$ area budget within a wearable SoC targeting 1\,mW average power. The 30\,s inference interval enables PE reuse to reduce footprint and local data reuse to limit memory traffic.
The developed CNN-LSTM accelerator (Fig.~\ref{fig:architecture}) consists of a controller, an output FIFO, and an interconnected 8-PE array. Each PE features a router, a register file, and a bit-flexible multiplier. Two of the PEs, designated as Super PEs, integrate enhanced functions, including pooling, activation, and quantization. This architectural partitioning simplifies control logic and minimizes hardware overhead.
The total on-chip memory capacity is 1.8~KB.

\begin{figure}[!b]
    \centering \vspace{-5mm}
    \includegraphics[width=.85\linewidth]{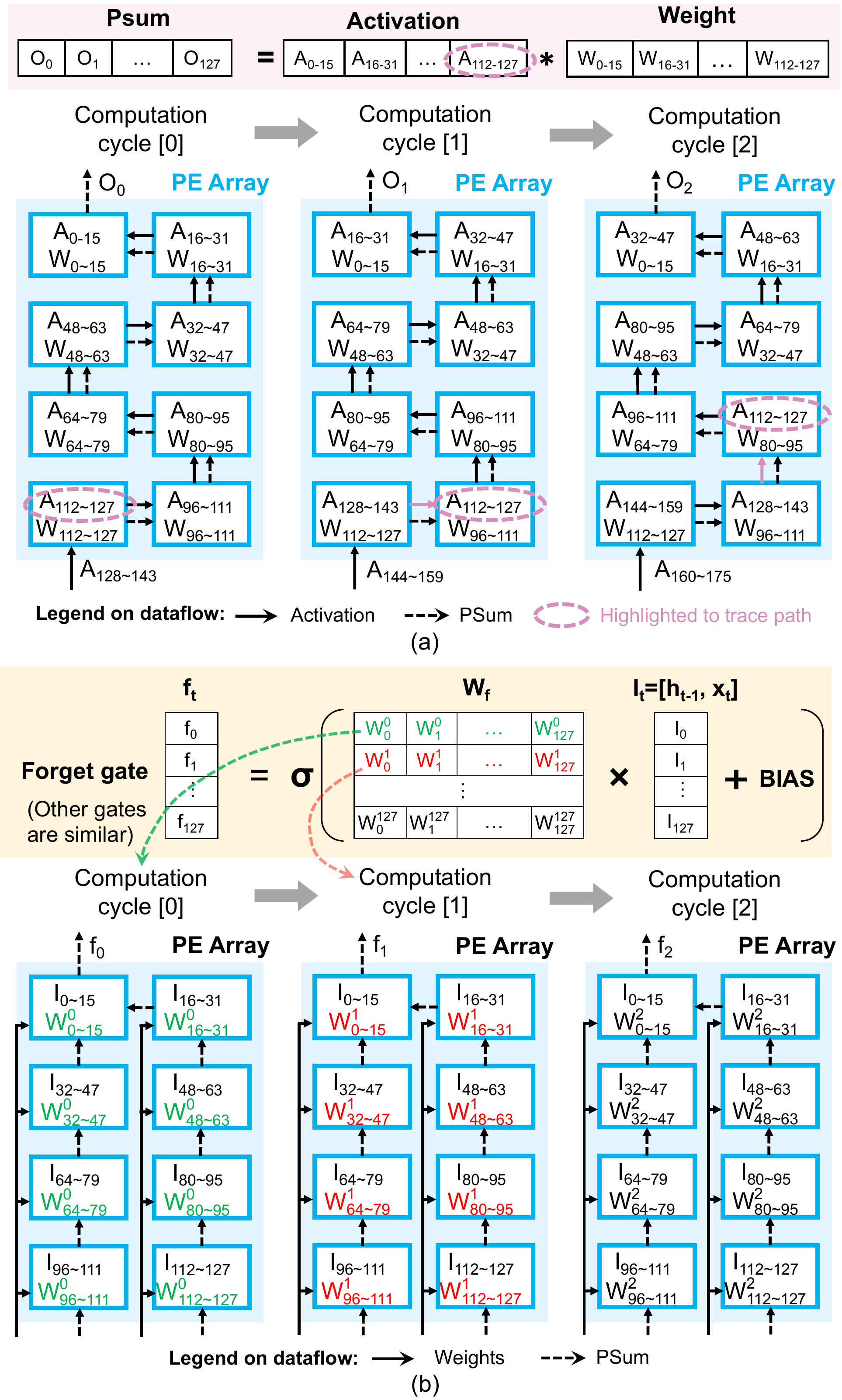} \vspace{-3pt} \vspace{-1mm}
    \caption{(a) Proposed serpentine dataflow for long-kernel CNNs, highlighting inter-PE activation reuse. (b) Input-stationary dataflow for LSTM computation, using the forget gate as an example.}
    \label{fig:data_flow}
\end{figure}

\section{Acceleration Techniques}

\subsection{Serpentine Dataflow for Long Kernels}

Memory accesses typically dominate accelerator energy consumption, making on-chip data reuse critical~\cite{chen2017eyeriss}. While prior weight- and output-stationary methods effectively maximize data reuse, they primarily target CNNs with small kernels (e.g., $3\times3$, $5\times5$) commonly found in computer vision tasks, providing poor support for the long temporal convolutions required in sleep staging and many EEG-based DL models.

To address this limitation, the proposed accelerator employs a serpentine dataflow, as illustrated in Fig.~\ref{fig:data_flow}(a). Rather than reloading overlapping activation windows, activations are propagated directly between neighboring PEs, enabling continuous reuse as they traverse the array. Consequently, only the new activations dictated by the convolution stride are fetched from SRAM during each computation cycle, while partial sums are accumulated along the PE chain until the final output is generated by the Super PE.

For short-kernel convolutions, a weight-stationary dataflow maximizes weight reuse. Conversely, LSTM layers utilize an input-stationary dataflow (Fig.~\ref{fig:data_flow}(b)) to efficiently share the input vector and hidden state across all four LSTM gates (forget, input, candidate, output)~\cite{xu2026inear}. In both modes, Super PEs handle partial-sum accumulation and final activation processing, minimizing memory accesses.

\subsection{Hybrid Static-Dynamic Mixed-Precision Quantization}

\begin{figure}[!b]
    \centering \vspace{-5mm}
    \includegraphics[width=1\linewidth]{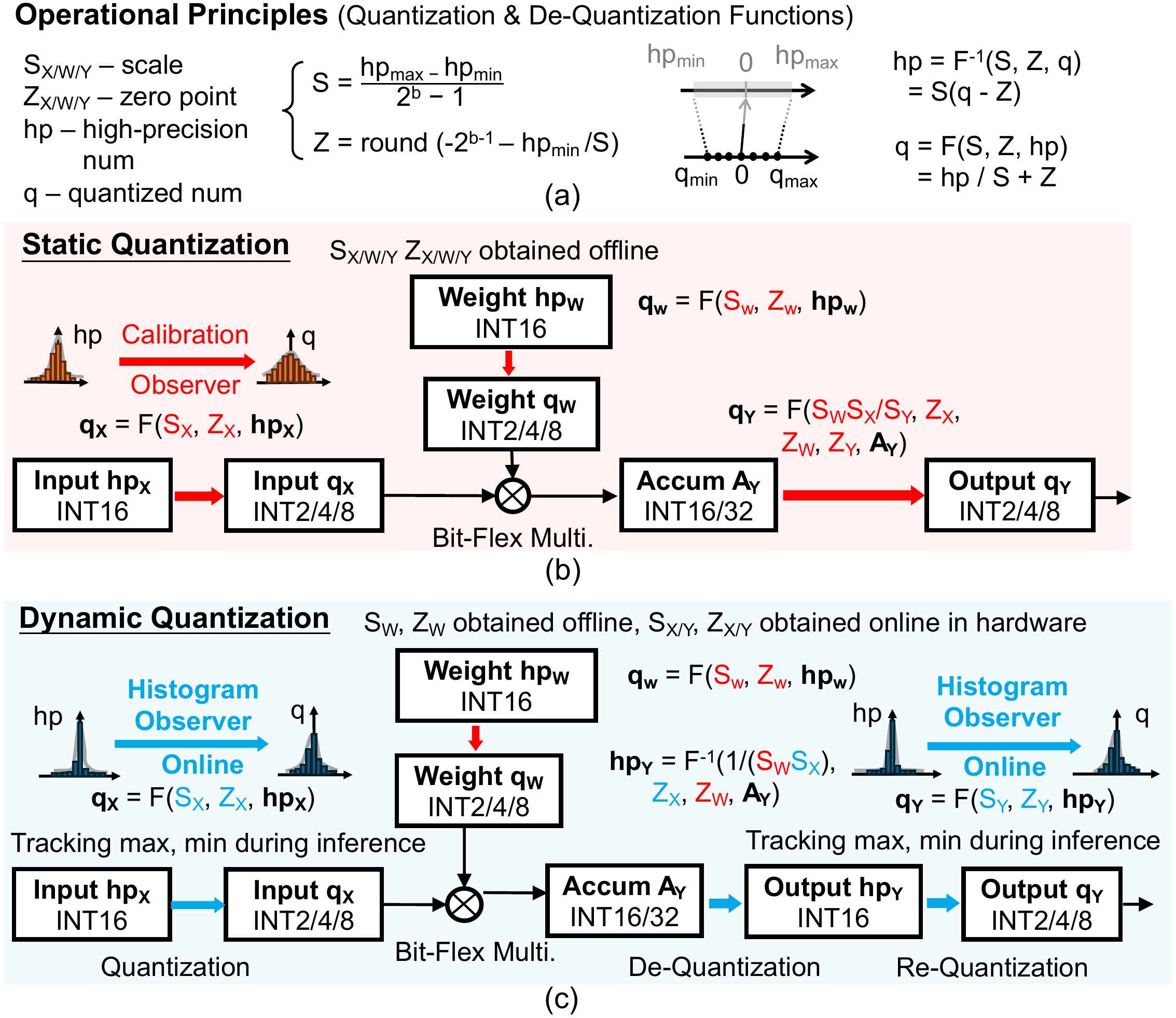} \vspace{-5mm} \vspace{-3pt}
    \caption{Proposed hybrid quantization scheme:
    (a) affine quantization and de-quantization mapping and key equations;
    (b) static quantization with offline-calibrated parameters for CNN layers; and
    (c) runtime dynamic quantization used for the LSTM and MLP layers, utilizing a histogram observer. The datapath supports INT2/4/8/16 operands with higher-precision accumulation.}
    \label{fig:quantization}
\end{figure}

Quantization reduces accelerator memory footprint and power.
However, conventional static quantization often degrades accuracy in recurrent networks since fixed activation ranges cannot adapt to time-varying distributions. To address this challenge, the proposed accelerator employs a hybrid static-dynamic mixed-precision quantization scheme. Static quantization is applied to layers with stable activation distributions, whereas runtime dynamic quantization adapts to layers with highly variable distributions. Furthermore, individual layers are assigned 2b, 4b, 8b, or 16b precisions based on their quantization sensitivity.

Fig.~\ref{fig:quantization}(a) shows the affine quantization and de-quantization mapping. Weights are quantized during training and stored in low-precision formats, using per-tensor scaling (or per-channel-/group-wise scaling for highly sensitive 2b layers). Activations utilize a per-tensor affine mapping. To prevent overflow during computation, partial sums are accumulated at higher precision.
In the static datapath (Fig.~\ref{fig:quantization}(b)), activation scales and zero-points are calibrated offline. High-precision accumulators are directly re-quantized to the output precision using a combined rescaling factor. This allows results to be written directly to memory without explicitly reconstructing intermediate values, minimizing inter-layer data traffic.
For dynamic quantization (Fig.~\ref{fig:quantization}(c)), the observer tracks each activation tensor's extrema range, then computes one scale and zero-point pair $(S,Z)$ before quantizing that tensor using the equations in Fig.~\ref{fig:quantization}(a). Parameters are updated at each MLP layer input and separately for $x_t$ and $h_{t-1}$ at every BiLSTM time step in each layer and direction. Observer state is reset for each inference, while CNN quantization parameters remain fixed after offline calibration.
To balance accuracy and hardware overhead, these two modes are assigned per layer: CNNs use static quantization for stable activations, while LSTM and MLP use dynamic quantization to track varying distributions and limit recurrent error propagation.

\vspace{-2mm}
\subsection{Bit-Flexible MAC and Memory Scheduling}

\begin{figure}[!b]
    \centering \vspace{-4mm}
    \includegraphics[width=.9\linewidth]{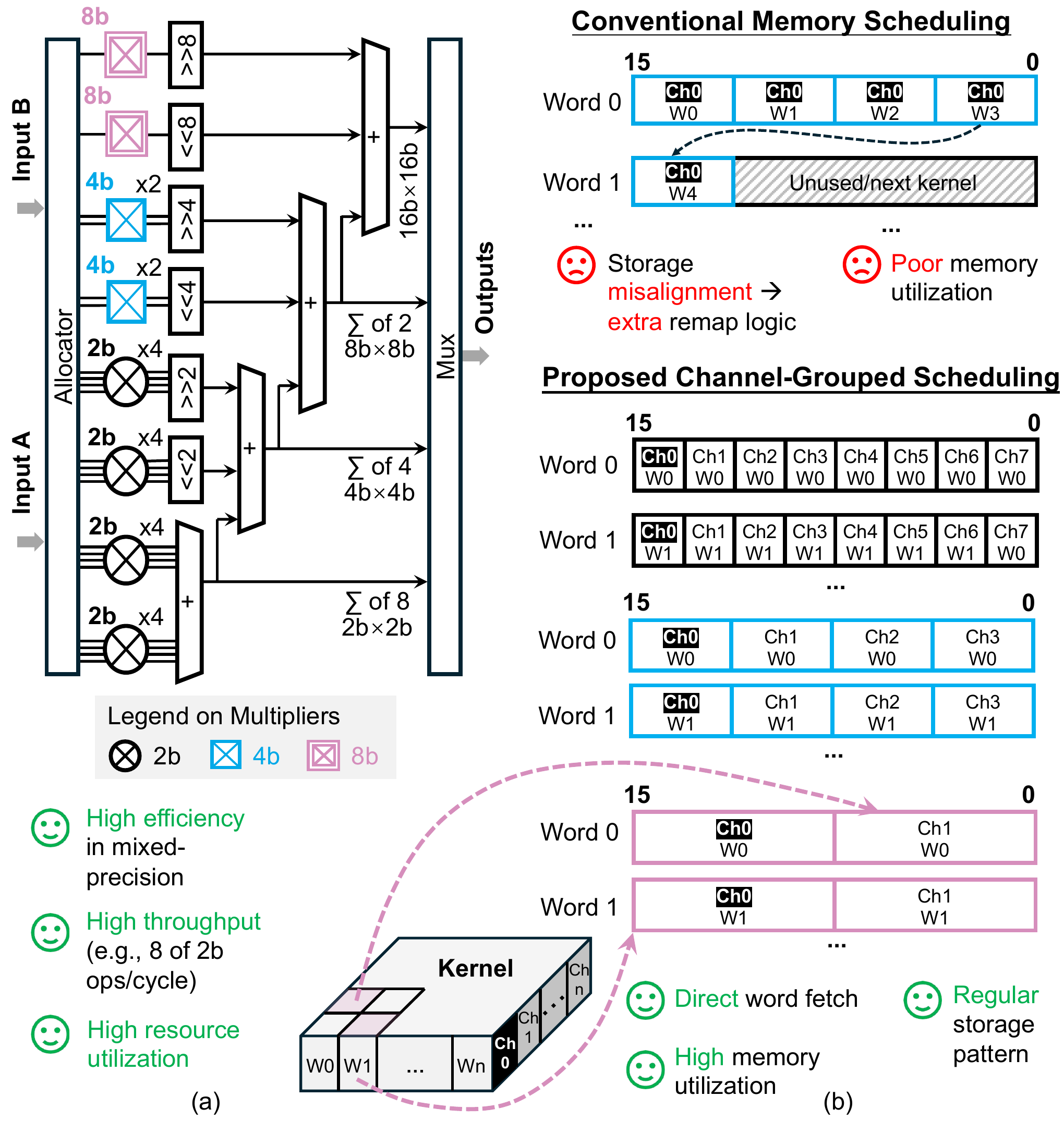} \vspace{-3mm}
    \caption{(a) Architecture of the developed bit-flexible multiplier. (b) Conventional and proposed channel-grouped memory scheduling.}
    \label{fig:bit_flexible}
\end{figure}

Prior work has explored hardware support for various quantization schemes~\cite{moon2024multipurpose_jssc}. However, to fully exploit the mixed-precision quantization scheme, the accelerator should support hardware-level mixed-precision MAC operations with dynamic quantization. Although various bit-flexible designs have been proposed~\cite{ryu2022bitblade,chen2024bbs}, they often incur area overhead, control complexity, or memory fragmentation from varying bit-widths.
To overcome these limitations, this work co-designs the bit-flexible MAC engine with memory scheduling. The proposed bit-flexible multiplier (Fig.~\ref{fig:bit_flexible}(a)) accepts two 16b inputs that are partitioned by an allocation network to execute eight 2b, four 4b, two 8b, or one 16b multiplication in parallel. The generated partial products are then shifted and accumulated to produce the final result. Compared with a state-of-the-art design~\cite{ryu2022bitblade}, this work delivers identical throughput while reducing the area (CMOS process scaled using~\cite{sarangi2021deepscaletool}) for 8b and 16b multiplications by 54.5\% and 77.3\%, respectively.

Variable bit widths cause memory fragmentation and complex alignment in conventional memory layouts. This work proposes channel-grouped memory scheduling (Fig.~\ref{fig:bit_flexible}(b)), which packs multiple channels into each 16b memory word according to operand precision.
Consequently, each access fetches a complete, aligned computation group, eliminating cross-word remapping and maximizing bandwidth utilization.

\section{Experimental Results}

\begin{figure}[!ht]
    \centering \vspace{-3mm}
    \includegraphics[width=.95\linewidth]{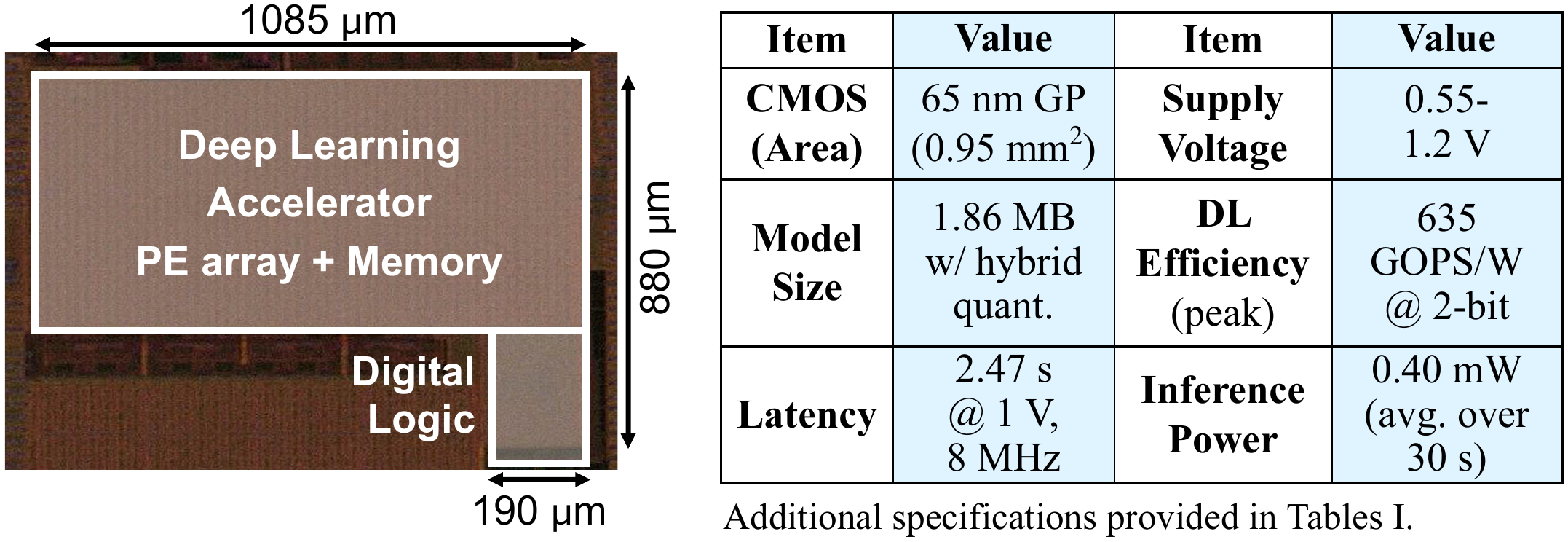} \vspace{-3pt} \vspace{-2mm}
    \caption{Micrograph of the fabricated SoC with key blocks highlighted and a summary of key chip specifications.}
    \label{fig:photo}
\end{figure}

A micrograph of the fabricated chip in 65\,nm CMOS and a summary of key chip specifications are shown in Fig.~\ref{fig:photo}.

\begin{figure}[!h]
    \centering \vspace{-3mm}
    \includegraphics[width=\linewidth]{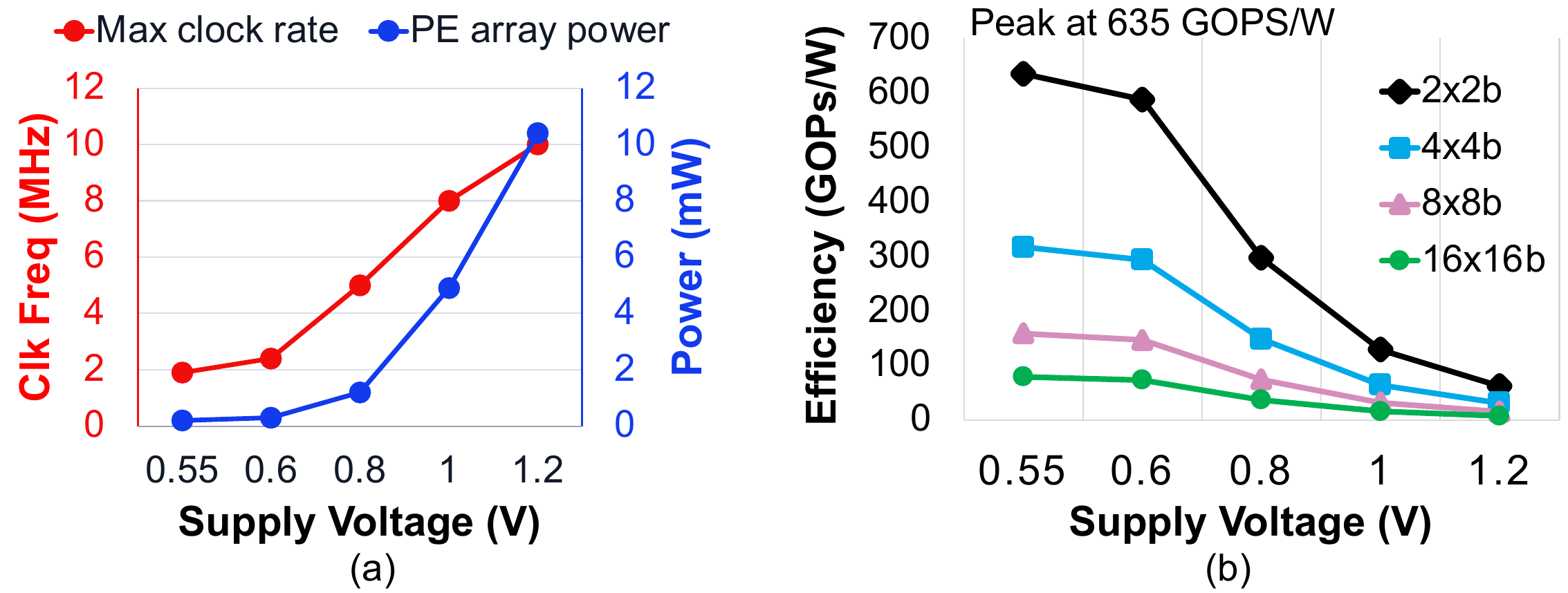} \vspace{-3pt} \vspace{-5mm}
    \caption{Voltage-scaling characteristics. (a) Maximum operating frequency and power consumption. (b) Peak power efficiencies at different precisions.}
    \label{fig:voltage_scaling}
\end{figure}

Fig.~\ref{fig:voltage_scaling} shows the measured voltage-scaling performance. Increasing the supply from 0.55\,V to 1.2\,V scales the maximum frequency from 2\,MHz to 10\,MHz and power from 0.24\,mW to 10.48\,mW (Fig.~\ref{fig:voltage_scaling}(a)). Evaluated at maximum frequencies across 2b to 16b modes, the power efficiency peaks at 635\,GOPS/W during 2b operation at 0.55\,V (Fig.~\ref{fig:voltage_scaling}(b)).
All reported power measurements include the power consumed by the on-chip memory and peripheral modules, which is also accounted for in the reported power efficiency.

\begin{figure}[!t]
    \centering
    \includegraphics[width=.95\linewidth]{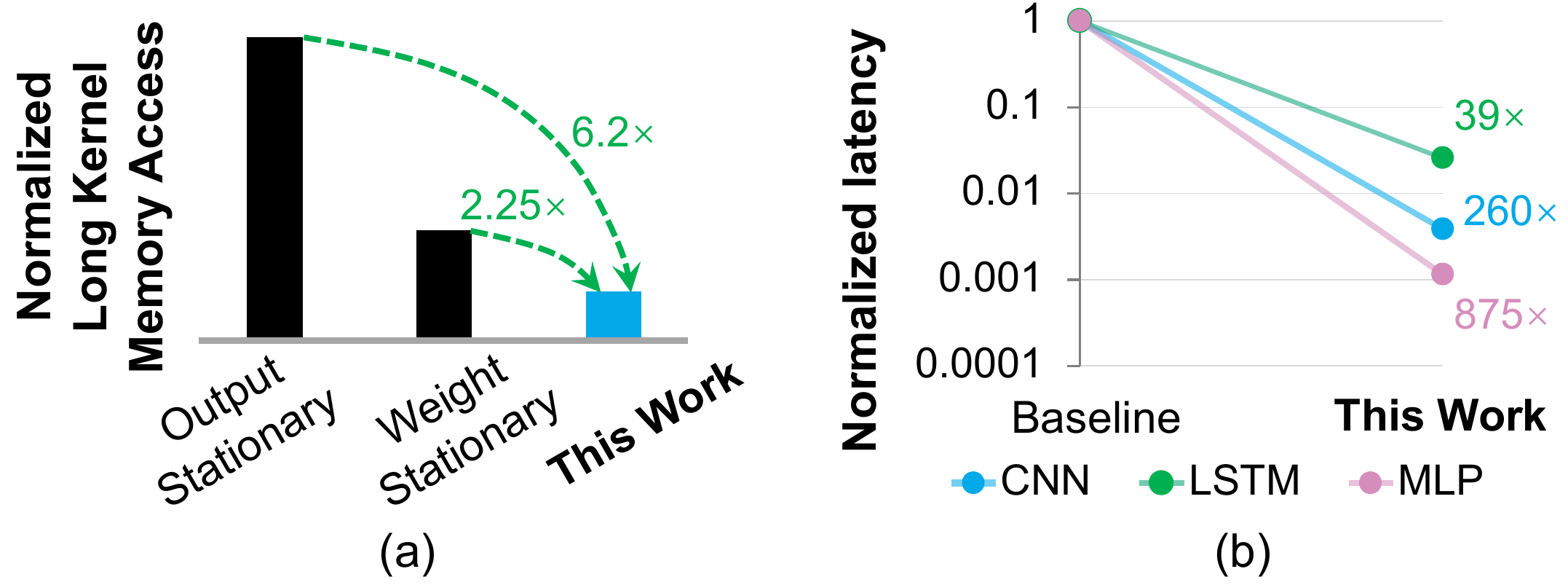} \vspace{-3pt} \vspace{-2mm}
    \caption{(a) Memory access reduction vs. mainstream dataflows for long-kernel convolutions. (b) Execution latency of the developed accelerator compared to a baseline 32b RISC-V processor across network layers.} \vspace{-2mm}
    \label{fig:memory_power}
\end{figure}

Next, we evaluated the memory movement efficiency of the proposed serpentine dataflow. For the long CNN kernels in our model, the proposed dataflow reduces SRAM accesses by 6.2$\times$ and 2.25$\times$ compared to conventional output- and weight-stationary dataflows on the same PE array (Fig.~\ref{fig:memory_power}(a)).
Benchmarked against an iso-frequency, single-core 32b \mbox{RISC-V} processor with a four-stage pipeline (execution latency obtained through cycle-accurate simulation), the proposed accelerator achieves speedups of 260$\times$, 39$\times$, and 875$\times$ for the CNN, LSTM, and MLP layers, respectively (Fig.~\ref{fig:memory_power}(b)), yielding an overall model speedup of 258$\times$.
Without a dedicated accelerator, inference would either fail to meet real-time requirements or require a high clock frequency with prohibitive power.

The conference version evaluated the model on two held-out subjects excluded from training~\cite{xu2026inear}. This work extends the evaluation to all 62 subjects in the MASS-SS3 dataset using 31-fold leave-two-subject-out cross-validation.
To assess both general sleep-staging and N3 detection for auditory stimulation, two model variants are reported: Ver-Gen (maximizing overall 5-class accuracy) and Ver-N3 (maximizing N3 sensitivity).
Table~\ref{tab:ml_model_comparison} compares this work with recent sleep-staging designs evaluated on MASS-SS3, highlighting its combination of ASIC implementation and hybrid static--dynamic quantization while achieving competitive accuracy. Numerical accuracy differences should be interpreted in the context of each study's input configuration, classification settings, and evaluation protocol.

\begin{table}[!t]
\centering
\caption{Key Model Performance Metrics and Comparison with Recent Sleep-Staging Designs Evaluated on MASS-SS3}
\vspace{-1mm}
\label{tab:ml_model_comparison}

\begingroup
\definecolor{paperbg}{RGB}{255,255,255}

\setlength{\tabcolsep}{1.5pt}
\renewcommand{\arraystretch}{1.08}
\fontsize{8.0pt}{8.6pt}\selectfont

\resizebox{\columnwidth}{!}{
\begin{NiceTabular}{|c|c|c|c|c|c|c|c|}
\CodeBefore
    \columncolor{thisworkblue}{3}
    \columncolor{thisworkblue}{4}
    \columncolor{paperbg}{5}
    \columncolor{paperbg}{6}
    \columncolor{paperbg}{7}
    \columncolor{paperbg}{8}
\Body
\hline
\Block{1-2}{\textbf{Publication}}
&
&
\Block{1-2}{\textbf{This work}}
&
&
\textbf{\makecell{OJCAS\\26~\cite{zhou2026fpga}}}
&
\textbf{\makecell{NEJ\\26~\cite{robitaille2026sleepvit}}}
&
\textbf{\makecell{Brain Sci.\\25~\cite{wijesinghe2025lightweight}}}
&
\textbf{\makecell{EUSIPCO\\24~\cite{seraphim2024structure}}}
\\
\hline

\Block{1-2}{\textbf{Hardware}}
&
&
\Block{1-2}{\textbf{ASIC}}
&
&
FPGA
&
\makecell{ASIC\\(sim.)}
&
No
&
No
\\
\hline

\Block{1-2}{\textbf{Input signal}}
&
&
\Block{1-2}{EEG $\times 2$}
&
&
EEG $\times 1$
&
EEG $\times 1$
&
EEG $\times 1$
&
EEG $\times 8$
\\
\hline

\Block{2-2}{\textbf{Architecture}}
&
&
\Block{1-2}{CNN-LSTM}
&
&
\Block{2-1}{\makecell{CNN-\\LSTM}}
&
\Block{2-1}{\makecell{Modified\\ViT}}
&
\Block{2-1}{\makecell{FNN +\\Trans. Vec.}}
&
\Block{2-1}{\makecell{2-stage\\Trans.}}
\\
\cline{3-4}
&
&
Ver-Gen&
Ver-N3&
&
&
&
\\
\hline

\Block{1-2}{\textbf{Quantization}}
&
&
\Block{1-2}{\textbf{\makecell{Yes \\2/4/8/16b Static+Dyn.}}}
&
&
\makecell{Yes\\16b Static}
&
\makecell{Yes\\8/16b Static}
&
No
&
No
\\
\hline

\Block{1-2}{\textbf{Accuracy}}
&
&
\textbf{\makecell{86.14\%\\(w/ Quant)}}
&
\makecell{84.87\%\\(w/ Quant)}
&
\makecell{83.7\%\\(w/ Quant)}
&
82.90\%
&
84.1\%
&
84.40\%
\\
\hline

\Block{3-1}{\textbf{\makecell{Deep\\sleep\\(N3)}}}
&
Sensitivity
&
\makecell{84.90\%\\(w/ Quant)}
&
\textbf{\makecell{95.75\%\\(w/ Quant)}}
&
--
&
--
&
--
&
--
\\
\cline{2-8}

&
\makecell{Per class\\F1 score}
&
\makecell{84.28\%\\(w/ Quant)}
&
\textbf{\makecell{84.56\%\\(w/ Quant)}}
&
--
&
\makecell{62.92\%\\(Est.)}
&
72.1\%
&
--
\\
\cline{2-8}

&
\makecell{Specificity}
&
\textbf{\makecell{98.10\%\\(w/ Quant)}}
&
\makecell{96.60\%\\(w/ Quant)}
&
--
&
--
&
--
&
--
\\
\hline
\end{NiceTabular}
}
\endgroup \vspace{-3mm}
\end{table}

\begin{figure}[!h]
    \centering \vspace{-2mm}
    \includegraphics[width=\linewidth]{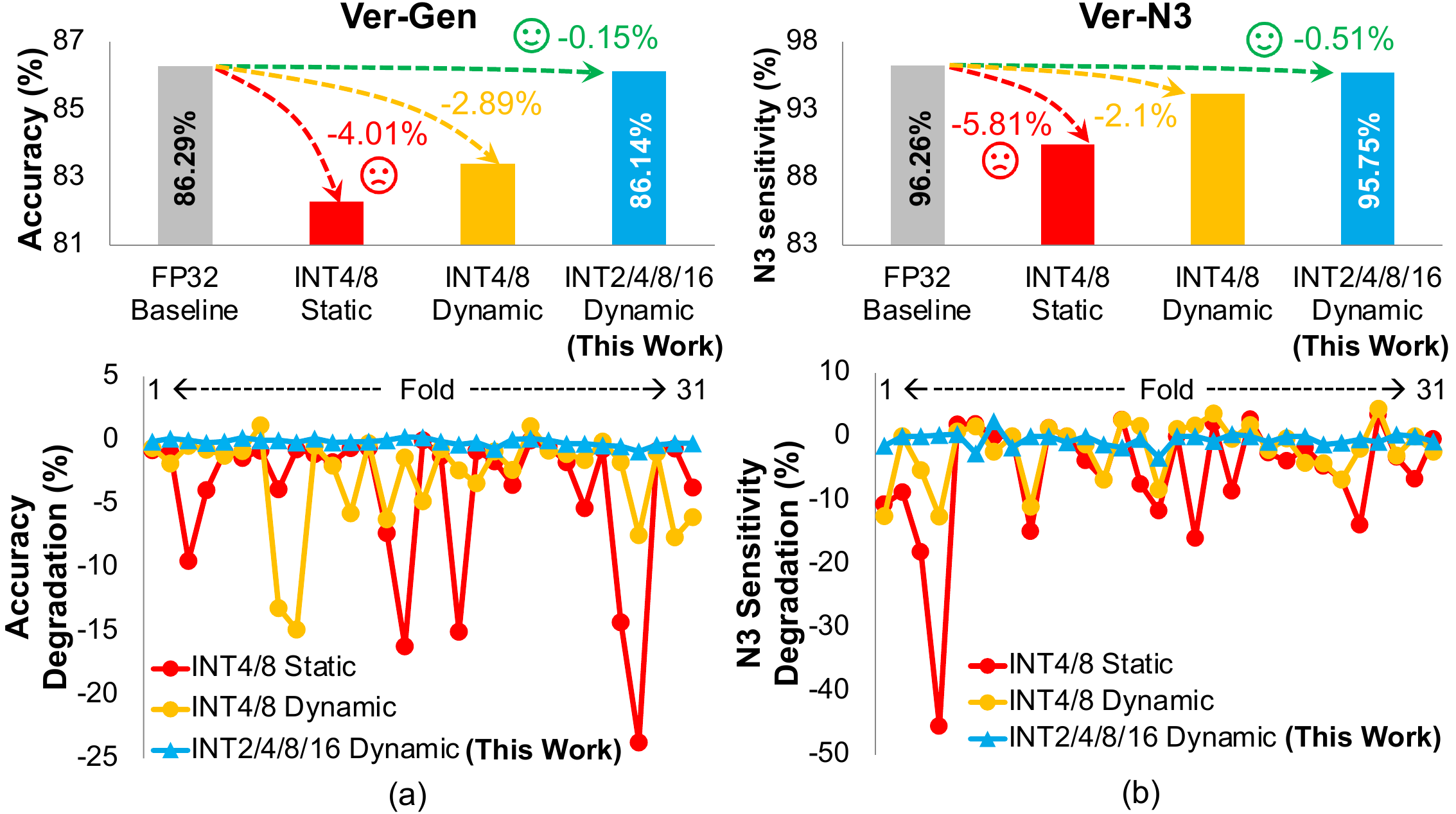} \vspace{-3pt} \vspace{-6mm}
    \caption{Experimental results of the mixed-precision hybrid quantization over 31 folds for (a) Ver-Gen and (b) Ver-N3, respectively. Conventional static quantization causes significant performance degradation, whereas the proposed mixed-precision dynamic quantization retains the accuracy and N3 sensitivity with a negligible degradation of 0.15\% and 0.51\%, respectively.} \vspace{-3mm}
    \label{fig:exp_quant}
\end{figure}

Fig.~\ref{fig:exp_quant} shows the experimental results of the mixed-precision hybrid quantization scheme compared with the floating-point (FP32) baseline. Applying full static INT4/8 quantization to the model degrades the accuracy of Ver-Gen by 4.01\% and the N3 sensitivity of Ver-N3 by 5.81\%, averaged across the 31 folds. Incorporating dynamic quantization at the same precision reduces the degradation to 2.89\% and 2.1\%, while the full INT2/4/8/16 mixed-precision dynamic quantization almost completely recovers the performance, leaving a negligible degradation of only 0.15\% and 0.51\%, respectively. The mixed-precision approach reduces the model size to below 1.86\,MB (4.16$\times$ reduction from baseline).
The 2b, 4b, 8b, and 16b groups account for approximately 0.007\%, 2.7\%, 90.8\%, and 6.5\% of the total operations, respectively, excluding max-pooling.
At 8\,MHz and a 1\,V supply, inference takes 2.47\,s, with an average power consumption of 0.40\,mW when performed once every 30\,s (matching the labeled epoch duration of the dataset).

\section{Conclusion}
This letter presents a 65\,nm mixed-precision CNN-LSTM accelerator for EEG-based sleep staging. A serpentine dataflow reduces long-kernel memory overhead, while hybrid mixed-precision quantization cuts model size by over 4$\times$ at negligible performance cost. Achieving a 258$\times$ speedup and N3 sensitivity of 95.75\%, it enables stage-specific sleep modulation, with techniques broadly applicable to EEG-based workloads. \vspace{-1mm}

\bibliographystyle{IEEEtran}
\bibliography{ref_SSCL}

\vfill

\end{document}